\documentclass[conference, a4paper]{IEEEtran}
\IEEEoverridecommandlockouts
\usepackage{cite}
\usepackage{amsmath,amssymb,amsfonts}
\usepackage{algorithmic}
\usepackage{graphicx}
\usepackage{textcomp}
\usepackage{xcolor}
\usepackage{caption}
\usepackage{subcaption}
\usepackage{url}
\usepackage{enumitem}
\def\BibTeX{{\rm B\kern-.05em{\sc i\kern-.025em b}\kern-.08em
    T\kern-.1667em\lower.7ex\hbox{E}\kern-.125emX}}
\begin{document}

\title{Resource-efficient DNNs for Keyword Spotting using Neural Architecture Search and Quantization}

\author{\IEEEauthorblockN{David Peter}
\IEEEauthorblockA{\textit{Signal Processing and} \\ \textit{Speech Communication Laboratory} \\
\textit{Graz University of Technology}\\
Graz, Austria \\
david.peter@student.tugraz.at}
\and
\IEEEauthorblockN{Wolfgang Roth}
\IEEEauthorblockA{\textit{Signal Processing and} \\ \textit{Speech Communication Laboratory} \\
\textit{Graz University of Technology}\\
Graz, Austria \\
roth@tugraz.at}
\and
\IEEEauthorblockN{Franz Pernkopf}
\IEEEauthorblockA{\textit{Signal Processing and} \\ \textit{Speech Communication Laboratory} \\
\textit{Graz University of Technology}\\
Graz, Austria \\
pernkopf@tugraz.at}
}

\maketitle

\begin{abstract}
This paper introduces neural architecture search (NAS) for the automatic discovery of small models for keyword spotting (KWS) in limited resource environments. We employ a differentiable NAS approach to optimize the structure of convolutional neural networks (CNNs) to maximize the classification accuracy while minimizing the number of operations per inference. Using NAS only, we were able to obtain a highly efficient model with 95.4\% accuracy on the Google speech commands dataset with 494.8 kB of memory usage and 19.6 million operations. Additionally, weight quantization is used to reduce the memory consumption even further. We show that weight quantization to low bit-widths (e.g. 1 bit) can be used without substantial loss in accuracy. By increasing the number of input features from 10 MFCC to 20 MFCC we were able to increase the accuracy to 96.3\% at 340.1 kB of memory usage and 27.1 million operations. 
\end{abstract}


\section{Introduction}
Automatic speech recognition (ASR) is becoming an increasingly important technology for user interaction with everyday consumer devices. However, ASR systems are typically complex and computation-intensive, i.e. running ASR in always-on mode results in a steady high energy consumption. This is especially problematic for mobile devices whose batteries are drained quickly when running ASR permanently. 

A common approach to this challenge is to run a low-cost keyword spotting (KWS) system that is listening permanently only for a limited set of prespecified keywords. Upon detection of such a keyword, a full ASR system is triggered which then listens for a rich set of user commands.

\noindent
The requirements of a KWS system are:
\begin{enumerate}[label=\roman*.]
	\item The system should be resource-efficient to mitigate the aforementioned energy problem,
	\item it should run in real-time, and
	\item it should be accurate to maintain a high user-experience.
\end{enumerate}

In this paper, we aim to find efficient and small KWS models for limited resource environments. We make use of neural architecture search (NAS) --- a popular technique to automate the design of deep neural network (DNN) architectures. A NAS algorithm tries to find the best performing neural network architecture within a given search space using some arbitrary search method. To obtain efficient DNN models we use the techniques from ProxylessNAS \cite{Cai2019}. However, unlike ProxylessNAS \cite{Cai2019} whose focus is on finding models for large-scale image classification tasks such as ImageNet, our focus lies on small-scale KWS tasks such as Google speech commands \cite{Warden2018}. 

We quantize model weights during NAS by default to 8 bits to better compare to \cite{Zhang2017}. In \cite{Zhang2017}, model weights are also quantized to 8 bits. Using NAS we obtain a model with 95.4\% accuracy on the test set of the Google speech commands dataset \cite{Warden2018} while using 494.8 kB of memory (for the storage of weights) and 19.6 million operations per inference. Once the efficient DNN architecture has been found, we compare different weight quantization schemes to show that the memory requirements can be reduced even further. Compressing the weights of this model to 1 bit allowed us to retain a test accuracy of 94.7\% while only using 61.85 kB of memory and 19.6 million operations.

Furthermore, we show that the number of features extracted from the raw audio waveform has a substantial impact on the performance. By changing the number of extracted MFCC features from 10 to 20 we were able to find a model with a test accuracy of 96.3\% using 340.1 kB of memory and 27.1 million operations. 

The outline of the paper is as follows: Section~\ref{sec:related_work} introduces the related work. In Section~\ref{sec:methods} we present our NAS configuration, the trade-off objective between accuracy and number of operations, and the weight quantization methods utilized in this paper. The experimental setup and the discussion of the results is shown in Section~\ref{sec:experiments}. Finally, Section~\ref{sec:conclusion} provides the conclusion. Code related to the paper is available online at \url{https://github.com/dapeter/nas-for-kws}.

\section{Related Work}
\label{sec:related_work}
Popular NAS approaches use concepts such as reinforcement learning \cite{Pham2018},
gradient based methods \cite{Liu2019} or even evolutionary methods \cite{Liu2018} for exploring the search space.
Recently, NAS techniques have also been used to find architectures that are specifically tailored to the underlying hardware \cite{Cai2019,Tan2019} by, for instance, additionally minimizing memory requirements, number of operations, or latency of the resulting model. Therefore, NAS techniques are well-suited for finding DNNs that run on mobile phones or even embedded devices.

Conventional NAS algorithms require thousands of architectures to be trained to find the final architecture. Not only is this very time consuming but in some cases even infeasible, especially for large-scale tasks such as ImageNet or when computational resources are limited. Therefore, several proposals have been made to reduce the computational overhead. So called \emph{proxy tasks} for reducing the search cost are training for fewer epochs, training on a smaller dataset or learning a smaller model that is then scaled up. As an example, \cite{Zoph2018} searches for the best convolutional layer on the CIFAR-10 dataset which is then stacked and applied to the much larger ImageNet dataset. Many other NAS methods implement this technique with great success \cite{Liu2018a,Real2019,Cai2018,Liu2019,Tan2019,Luo2018}. Another approach, called EfficientNet \cite{Tan2019a}, employs the NAS approach from \cite{Tan2019} to find a small baseline model which is subsequently scaled up. By scaling the three dimensions depth, width and resolution of the small baseline model, they obtain state-of-the-art performance.

Although model scaling and stacking achieves good performances, the convolutional layers optimized on the proxy task may not be optimal for the target task. Therefore, several approaches have been proposed to get rid of proxy tasks. Instead, architectures are directly trained on the target task and optimized for the hardware at hand. In \cite{Cai2019} an overparameterized network with multiple parallel candidate operations per layer is used as the base model. Every candidate operation is gated by a binary gate which either prunes or keeps the operation. During architecture search, the binary gates are then trained such that only one operation per layer remains and the targeted memory and latency requirements are met. This is extended in \cite{Stamoulis2019} by using shared parameters among individual layers to avoid excessive parameter overhead during training.

In KWS, models using neural networks for keyword classification have become increasingly popular. They are not only easy to train, but they also achieve state-of-the-art performances on image and speech classification tasks. In \cite{Zhang2017} several models from the literature \cite{Chen2014,Sainath2015,Arik2017,Sun2016} are evaluated on the Google speech commands dataset \cite{Warden2018}. They compare their models in terms of accuracy, memory requirements and number of operations. To allow easy deployment on microcontrollers they train their models using 32 bit float numbers and quantize the weights after training to 8 bit fixed-point numbers. They argue that fixed point numbers have been shown to suffice to run neural networks with minimal loss in accuracy \cite{Suda2016,Qiu2016,Lai2017}. We use the results from \cite{Zhang2017} in our experiments as a baseline for our models.

In binarized neural networks (BNNs) the resolution of both weights and activations is reduced to binary values $\{-1,1\}$ \cite{Hubara2016} while still retaining state-of-the art performance on image classification tasks such as MNIST, CIFAR-10 and SVHN. BNNs rely on the straight-through estimator (STE) \cite{Hinton2012,Bengio2013} which approximates the gradient of non-differentiable or piecewise constant functions, such as quantizers, by the non-zero gradient of some other function. The STE was also used in \cite{Hubara2016} to substitute the derivative of piecewise-constant functions. Another method for quantizing neural networks involves a Bayesian approach to learn weight distributions over discrete weights \cite{Roth2019}. The discrete-weight network is then obtained by selecting the most probable weights. For a comprehensive overview of resource-efficient methods for DNNs, we refer the interested reader to \cite{Roth2020}.

\section{Methods}
\label{sec:methods}
\subsection{Neural Architecture Search}
We use convolutional neural networks (CNNs) for keyword classification. Our goal is to find well performing architectures for different computing regimes. To achieve this, we use the multi-objective NAS approach from \cite{Cai2019} called ProxylessNAS. In ProxylessNAS, hardware-aware optimization is performed by NAS to optimize the model accuracy and latency on different hardware platforms. However, we optimize for accuracy and number of operations and not for latency. By optimizing for number of operations, the model size is implicitly optimized as well. The trade-off between the accuracy and the number of operations is established by regularizing the architecture loss as explained in Section~\ref{sec:tradeoff}.

ProxylessNAS constructs an overparameterized network with multiple parallel candidate operations per layer as the base model. A single operation is denoted by $o_i$. Every operation is assigned a real-valued architecture parameter $\alpha_i$. The $N$ architecture parameters are transformed to probability values $p_i$ by applying the softmax function
\begin{equation}
	p_i = \frac{\exp(\alpha_i)}{\sum_{j}\exp(\alpha_j)}.
\end{equation}
Every candidate operation is gated by a binary gate $g_i$ which either prunes or keeps the operation. Only one gate per layer is active at a time. Gates are sampled randomly according to 
\begin{equation}
\label{eq:gates}
	[g_1, g_2, \dots, g_N] = \begin{cases}
	[1, 0, \dots, 0]       & \text{with prob. } p_1, \\
	[0, 1, \dots, 0]       & \text{with prob. } p_2, \\
	\,\quad \quad \dots \\
	[0, 0, \dots, 1]       & \text{with prob. } p_N. \\
	\end{cases}.
\end{equation}
Based on the binary gates and the layer input $x$, the output of the mixed operation $m_\mathcal{O}^\mathrm{Binary}$ (i.e. the output of the layer) is defined as
\begin{equation}
	m_\mathcal{O}^\mathrm{Binary} = \sum_{i=1}^{N} g_i o_i(x).
\end{equation}
During architecture search, the training of the architecture parameters $\alpha_i$ and the weight parameters of the operations $o_i$ are performed in an alternating manner. When training weight parameters, the architecture parameters are frozen and binary gates are sampled according to (\ref{eq:gates}). Weight parameters are then updated via standard gradient descent on the training set. When training architecture parameters, the weight parameters are frozen and the architecture parameters are updated on the validation set. Updating the architecture parameters via backpropagation requires computing $\partial L / \partial \alpha_i$ which is not defined due to the sampling in (\ref{eq:gates}). ProxylessNAS proposes two methods for estimating the gradient. Method one estimates the gradient of $\partial L / \partial p_i$ as $\partial L / \partial g_i$. This estimation is an application of the STE explained in Section~\ref{sec:weight_quantization}. With this estimation, backpropagation can be used to compute $\partial L / \partial \alpha_i$ since the gates $g_i$ are part of the computation graph and thus $\partial L / \partial g_i$ can be calculated using backpropagation. Method two utilizes REINFORCE \cite{Williams1992} to find the optimal binary gates that maximize a certain reward $R(\cdot)$. The reward $R(\cdot)$ is directly proportional to the model accuracy and inversely proportional to the number of operations of the model. We did not observe any benefits of using REINFORCE over the gradient based approach. Therefore, we use the gradient based approach in our experiments to approximate the gradient.

Our overparameterized network, shown in Table~\ref{model_structure}, consist of three stages, (i) an input stage, (ii) an intermediate stage, and (iii) an output stage. Stages (i) and (iii) are fixed to a 5$\times$11 convolution and a 1$\times$1 convolution respectively. The network size is lower bounded by the size of stage (i) and stage (iii), which however is negligible regarding the overall model size. We apply batch normalization \cite{Ioffe2015} followed by the ReLU \cite{Nair2010} non-linearity as an activation function after the convolutions of stages (i) and (iii). We use mobile inverted bottleneck convolutions (MBCs) \cite{Sandler2018} as our main building blocks in stage (ii).

Only convolutions from stage (ii) are optimized. MBCs have two learnable parameters, the expansion rate $e$ and the size $k$ of the quadratic $k$$\times$$k$ convolution kernel. MBCs consist of three separate convolutions, one 1$\times$1 convolution followed by a depthwise-separable 3$\times$3 convolution followed again by a 1$\times$1 convolution. The first two convolutions use batch normalization and ReLU activation functions. The third convolution only uses batch normalization. The first and third 1$\times$1 convolution change the number of feature maps by the expansion rate factor of $e$ and $1/e$ respectively. Stride (as stated in Table~\ref{model_structure}) is only applied to the first convolution of each stage. The stride can be different along certain dimensions. Therefore it is parametrized by (H, W), where H is the stride along the height and W is the stride along the width of a certain feature map.

During NAS, we allow MBCs with expansion rates $e \in \{1,2,3,4,5,6\}$ and kernel sizes $k \in \{3,5,7\}$ for selection. We also include the zero operation. Therefore, our overparameterized network has $\#e\cdot\#k + 1=19$ binary gates per layer. For blocks where the input feature map size is equal to the output feature map size we include skip connections. If a zero operation is selected as a block by NAS, the skip connection allows this particular layer to be skipped similar to \cite{Cai2019} resulting in an identity layer.

\begin{table}[t!]
	\caption{Three stage model used for the KWS task. Stages (i) and (iii) are fixed to a 5$\times$11 convolution and a 1$\times$1 convolution respectively. Stage (ii) convolutions are optimized.}
	\begin{center}
		\begin{tabular}{|c|c|c|c|c|}
			\hline
			Stage & Operation & Stride (H, W) & \#Channels & \#Layers \\
			\hline\hline
			(i) & Conv, 5$\times$11 & 1, 2 & 72 & 1 \\
			\hline
			(ii) & MBC[e], [k]$\times$[k] & 2, 2 & 72 & 12 \\
			& or Identity & & & \\
			\hline
			(iii) & Conv, 1$\times$1 & 1, 1 & 144 & 1 \\
			& Global Avg. Pooling & & &\\
			& Fully connected & & &\\
			\hline
		\end{tabular}
	\end{center}
	\label{model_structure}
\end{table}

\subsection{Accuracy and Operations Trade-off Objective}
\label{sec:tradeoff}
The trade-off between the network accuracy and the number of operations is established by regularizing the architecture loss similar as in \cite{Cai2019} according to 
\begin{equation}
\label{eq:regularization}
\mathrm{loss}_{arch} = \mathrm{CE}_{loss} \cdot \left(\frac{\log(\mathrm{ops}_{exp})}{\log(\mathrm{ops}_{target})}\right)^\beta
\end{equation} 
where $\mathrm{CE}_{loss}$ is the cross entropy loss, $\mathrm{ops}_{exp}$ the expected number of operations, $\mathrm{ops}_{target}$ the target number of operations and $\beta$ the regularization parameter. The expected number of operations $\mathrm{ops}_{exp}$ is the overall number of operations expected to be used in the convolutions and in the fully connected layer based on the probabilities $p_i$ of the final network. For stages (i) and (iii) we simply sum the number of operations used in the convolutions and in the fully connected layer. For stage (ii) layers, the expected number of operations is defined as the weighted sum of the numbers of operations used in $o_i$ weighted by the probabilities $p_i$. 

Establishing an accuracy/operations trade-off using (\ref{eq:regularization}) is usually achieved by fixing $\mathrm{ops}_{target}$ and choosing $\beta$ such that the number of operations of the final architecture after NAS are close to $\mathrm{ops}_{target}$. If $\mathrm{ops}_{target}$ is close to $\mathrm{ops}_{exp}$, the right term of (\ref{eq:regularization}) becomes one thus leaving the $\mathrm{CE}_{loss}$ unchanged. However, if $\mathrm{ops}_{exp}$ is larger or smaller than $\mathrm{ops}_{target}$, the $\mathrm{CE}_{loss}$ is scaled up or down respectively.  

In our work, we focus on obtaining a wide variety of networks with a different number of operations, rather than reaching a certain number of target operations. Therefore, we kept $\mathrm{ops}_{target}$ fixed while varying $\beta$. Varying $\beta$ while fixing $\mathrm{ops}_{target}$ resulted in more diverse networks in terms of number of operations than fixing $\beta$ and varying $\mathrm{ops}_{target}$.

\subsection{Weight Quantization}
\label{sec:weight_quantization}
Our goal is to find resource efficient models for limited resource environments. Therefore, we use weight quantization to reduce the model size in terms of memory consumption even further. Unless noted otherwise, we quantize our model weights to 8 bits during NAS. This allows us to compare our models to the quantized 8 bit models in \cite{Zhang2017} while also reducing the model size substantially.

We perform quantization similar to \textit{quantize\textsubscript{k}} in \cite{Zhou2016}.  \textit{quantize\textsubscript{k}} takes a real-valued input $r_i \in [0, 1]$ and quantizes it to a $k$ bit number $r_o \in [0, 1]$. However, we want to quantize real-valued weights $w \in \mathbb{R}$ to $k$ bit weights $w_q \in [-1, 1]$. We achieve this by extending \textit{quantize\textsubscript{k}} to
\begin{equation}
\label{eq:quantizer}
	w_q = 2 \cdot \left[\frac{1}{2^k - 1} \mathrm{round}\left((2^k - 1) \frac{w + 1}{2} \right)\right]_{0}^1 - 1
\end{equation}
where $[x]_a^b$ is the clamping function
\begin{equation}
	[x]_a^b = \max(a, \min(x, b)) \in [a, b].
\end{equation}
The real-valued weights are quantized to a set of discrete values $\{q_1, q_2, \dots, q_N\}$. The quantizer is uniform since the quantization steps $q_{i+1} - q_i$ are equal. Non-uniform quantization is used by some methods by applying a linear quantization to the logarithm of the input \cite{Miyashita2016}. Furthermore, our quantizer is a Mid-Rise type quantizer which means that the origin is not included in the set of discrete output values. In contrast, a Mid-Tread type quantizer includes the origin in the set of discrete output values.

We compare two techniques for quantization: bit-rounding as a post-processing step and quantization aware training using the STE. Both techniques use (\ref{eq:quantizer}) to quantize the weights between +1 and -1.

Bit-rounding as a post-processing step is performed after training by simply quantizing the weights according to (\ref{eq:quantizer}). The STE is used to quantize the weights during training. The STE approximates the gradient of functions that are non-differentiable or whose gradient is zero almost everywhere such as quantization functions. We apply the quantizer from (\ref{eq:quantizer}) to quantize the weights during the forward pass. During backpropagation, the zero-gradient of the quantizer is replaced by the non-zero gradient of the linear function $f(x) = x$, i.e., $f'(x)=1$, which essentially passes the gradient through the quantizer. This allows us to train the network while simultaneously quantizing the weights.

\begin{figure}[t!]
	\centerline{\includegraphics[width=\linewidth]{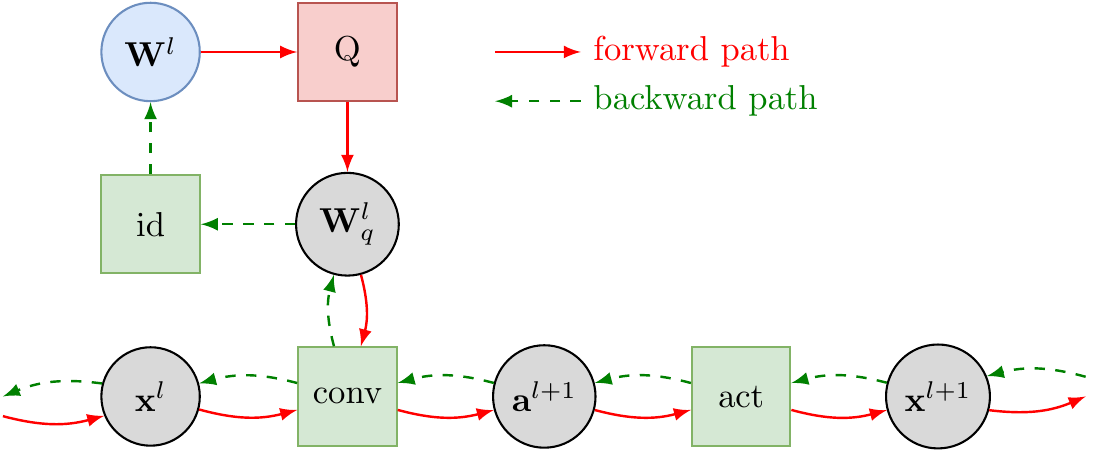}}
	\caption{Weight quantization using the STE on a typical convolutional layer (without batch normalization). Red boxes have zero gradients whereas green boxes have non-zero gradients. Weight updates are performed to the blue circle. During the forward pass, the weight tensor $\mathbf{W}^l$ is quantized to obtain the $k$ bit weight tensor $\mathbf{W}^l_q$ used in the convolution. The activation function is then applied to the output of the convolution $\mathbf{a}^{l+1}$ to obtain $\mathbf{x}^{l+1}$ which is the input tensor of the subsequent layer. During the backward pass, the STE replaces the derivative of the quantizer by the derivative of the identity function \textit{id}. Note that the activations are not quantized.} 
	\label{fig:ste}
\end{figure}

Figure~\ref{fig:ste} illustrates weight quantization using the STE on a typical convolutional layer (without batch normalization). During the forward pass, the quantizer $Q$ performs the quantization of the weight tensor $\mathbf{W}^l$ according to (\ref{eq:quantizer}). During the backward pass, the STE replaces the derivative of the quantizer $Q$ by the derivative of the identity function. In this way, the gradient can propagate back to $\mathbf{W}^l$.

Generally, we observe that quantization aware training with the STE performs better than bit-rounding as a post-processing step (cf. Section~\ref{sec:quantization}).

\section{Experiments}
\label{sec:experiments}
\subsection{Dataset}
We use the first version of the Google speech commands dataset \cite{Warden2018}. It consists of 65.000 1-second long audio files sampled with 16 bit at 16 kHz sampling frequency. Every audio file contains one utterance of an english word spoken by one person. In total there are 30 words spoken by thousands of people. The words are grouped into 30 different classes. We follow the procedure of \cite{Zhang2017} and use the following 10 classes ”Yes”, ”No”, ”Up”, ”Down”, ”Left”, ”Right”, ”On”, ”Off”, ”Stop” and ”Go” from the dataset. Likewise, we also include an "unknown" class which is a blend of randomly selected samples from the remaining 20 classes. Furthermore, a "silence" class is added. The "silence" class is artificially generated and consists of 1-second audio files containing a random slice of audio from a randomly selected noise sample provided by the Google Speech commands dataset. We made sure that the number of samples in the "unknown" class and in the "silence" class is equal to the average number of samples in all 30 classes to obtain equal class sizes. Our test set consists of the keyword samples from the test set and includes the same random selection of samples from the "unknown" and the "silence" class as used in \cite{Zhang2017} to make our results comparable.

\subsection{Experimental Setup}
We extract MFCC features from the raw audio streams before performing classification. Unless noted otherwise, we extract 10 MFCC features per 40ms frame. We use a stride length of 20ms which yields a total of 10$\times$51 features per 1 second of audio.

We augment all training samples with a random time shift of up to 100ms. Furthermore, 80\% of the training data is augmented with background noise. We use the noise samples from the Google speech commands dataset. It includes environmental sounds, speech, as well as white and pink noise. If a sample is augmented, it is first time shifted and then, in 80\% of the cases, overlayed with a 1 second interval of a random noise sample. The desired signal and the noise signal are mixed in a $(1-\epsilon):\epsilon$ ratio using $\epsilon \sim \mathcal{U}(0, 0.1)$.

Before actually performing the architecture search we pretrain the architecture blocks for 40 epochs at a learning rate of 0.05. During pretraining, architecture blocks are randomly sampled and trained on one batch of the training set. Sampling and training repeats until 40 epochs of training are passed. Optimization is performed with stochastic gradient descent using a mini-batch size of 100. After pretraining we perform an architecture search for 120 epochs with an initial learning rate of 0.2. The learning rate is decayed according to a cosine schedule \cite{Cai2019}. 

An architecture is trained until convergence after selection by the NAS procedure to optimize the performance. We use the same hyperparameters as in the architecture search process. 

\subsection{Efficient Architecture Search}
\label{sec:efficient_arch_search}
We search for architectures of different sizes by varying the regularization parameter $\beta \in \{0, 1, 2, 4, 8, 16\}$ to obtain a trade-off between the number of operations and accuracy. Furthermore, we kept $\mathrm{ops}_{target}$ fixed at $20 \cdot 10^6$ during all experiments. By default, we perform quantization aware training using the STE during NAS to quantize the model weights to 8 bits.

Our models have the same number of channels across all layers except for the last 1$\times$1 convolution where the number of feature maps is doubled with respect to the number of feature maps in the previous layers. We use networks with 72 channels in stage (i) and (ii) and 144 channels in stage (iii) as our base networks. To achieve more diversity, we employ a channel multiplier $\omega$ which scales the number of channels in each layer. We evaluated $\omega \in \{0.75, 1, 1.25\}$ and rounded the resulting number of channels to the closest multiple of 8 \cite{Sandler2018}.

\begin{figure}[t!]
	\centerline{\includegraphics[width=0.85\linewidth]{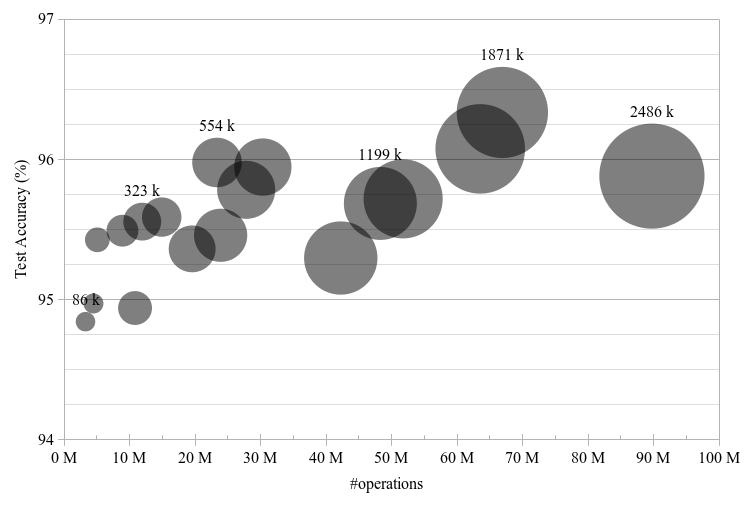}}
	\caption{Test accuracy versus number of operations using multi-objective NAS for varying trade-offs $\beta$. The model size (i.e. memory used to store the weights in bytes) corresponds to the circle size. Numbers above circles represent the model size in bytes for some selected models.}
	\label{fig:network_performances}
\end{figure}

Figure~\ref{fig:network_performances} shows the test accuracy versus the number of operations for all models found using NAS. The model size (i.e. memory used to store weights in bytes) corresponds to the circle size. We observe a variety of models from different operation regimes.

\begin{table}[t!]
	\caption{Test accuracy, number of operations and memory requirements of three models obtained using NAS.}
	\begin{center}
		\begin{tabular}{|c|c|c|c|c|}
			\hline
			Architecture & Test Acc. (\%) & Operations & Memory \\\hline\hline
			Hello Edge DS-CNN \cite{Zhang2017} & 94.4 & 5.4 M & 38.6 kB \\\hline
			Hello Edge DS-CNN \cite{Zhang2017} & 94.9 & 19.8 M & 189.2 kB \\\hline
			Hello Edge DS-CNN \cite{Zhang2017} & 95.4 & 56.9 M & 497.6 kB \\\hline\hline
			Ours (Fig.~\ref{fig:models}c), $\omega=0.75$ & 95.0 & 4.6 M & 89.8 kB \\\hline
			Ours (Fig.~\ref{fig:models}b), $\omega=1.25$ & 95.4 & 19.6 M & 494.8 kB \\\hline
			Ours (Fig.~\ref{fig:models}a), $\omega=1$ & 96.0 & 23.4 M & 554.0 kB \\\hline
		\end{tabular}
	\end{center}
	\label{tab:model_comparison}
\end{table}

\begin{figure}[t!]
	\centerline{\includegraphics[width=0.9\linewidth]{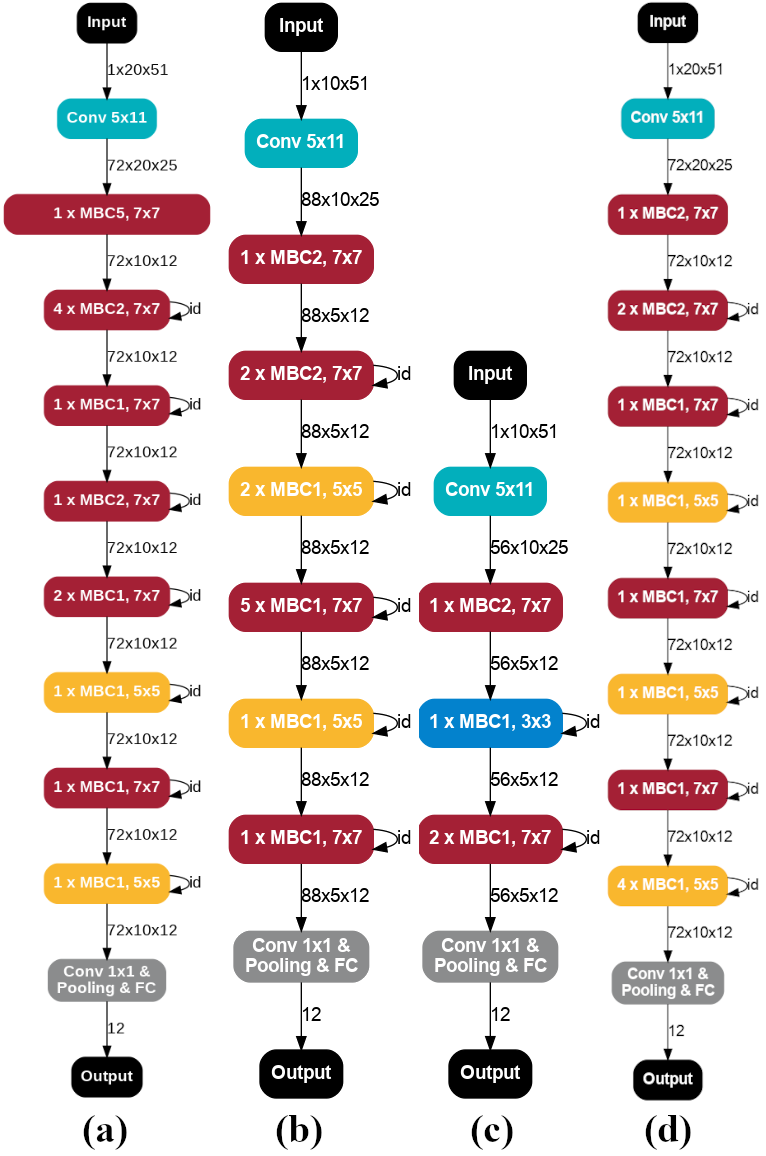}}
	\caption{Large (a), medium (b) and small (c) sized models found by NAS using 10 MFCCs (cf. Section~\ref{sec:efficient_arch_search}). Identical consecutive MBC blocks are depicted as one block for representational purposes. Model (d) is the best model found by NAS among all models using 20 MFCCs (cf. Section~\ref{sec:mfcc_features}).}
	\label{fig:models}
\end{figure}

We compare three of our models with depthwise-separable convolutional neural network (DS-CNN) models from \cite{Zhang2017} in Table~\ref{tab:model_comparison}. Figure~\ref{fig:models} (a), (b) and (c) provide the corresponding network topologies. Inverted bottleneck convolutions are denoted by \emph{MBC[e], [k]$\times$[k]} with expansion rate $e$ and kernel size $k$. Expansion rates,  kernel sizes and number of layers are determined by NAS. MFCC features at the input are processed first by a 5$\times$11 convolution which reduces the width by a factor of 2 by means of a strided convolution. The first inverted bottleneck convolution then reduces the height and width by a factor of 2 by means of a strided convolution. The last stage contains a 1$\times$1 convolution as well as global average pooling followed by a fully connected layer. Individual MBC blocks have skip connections if the output of the block has the same shape as the input of the block. Skip connections are drawn symbolically and are denoted by \emph{id}. The output of the convolutions are three dimensional tensors, with dimensions C$\times$H$\times$W, where C is the number of channels, H is the height and W is the width of the corresponding tensor.

DS-CNN models are currently among the best performing models. Still, NAS allows us to find models that match or even outperform the baseline  while requiring substantially fewer operations.

\subsection{Weight Quantization}
\label{sec:quantization}
Reducing the model size in terms of memory consumption is not only possible by efficient architectures but also by efficient weight representations. Several methods are known, including weight pruning and weight quantization \cite{Roth2020}. We apply weight quantization to reduce our model sizes even further.

We perform several tests where we quantize the weight tensors of our second model from Table~\ref{tab:model_comparison} in the range of 8 bit to 1 bit. Note that our models use 8 bit weights by default. The activations are kept at full precision. We investigate the influence of weight quantization on the test accuracy. We compare two different methods for weight quantization as discussed in Section~\ref{sec:weight_quantization}:
\begin{itemize}
	\item Quantization aware training using the STE.
	\item Quantization as a post-processing step by rounding parameters of a trained network to a prespecified parameter bit-width.
\end{itemize}

\begin{figure}[t!]
	\centerline{\includegraphics[width=0.95\linewidth]{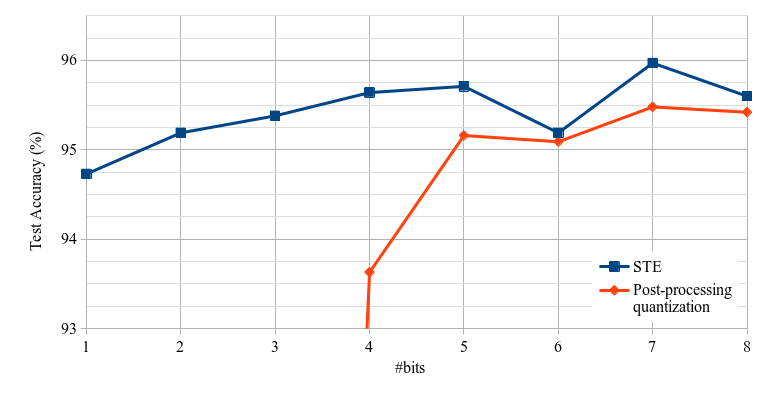}}
	\caption{Test accuracy versus number of bits for two weight quantization methods evaluated on our second model from Table~\ref{tab:model_comparison} for bit-widths in the range of $\{1, 2, \dots,8\}$. For every bit-width, the model is trained from scratch and quantized according to the quantization method. Results for STE (blue) are obtained by learning quantized weights during training. Results for post-processing quantization (red) are obtained by rounding the weights after training.}
	\label{fig:quantization_comparison}
\end{figure}

Figure~\ref{fig:quantization_comparison} includes the results and compares the test accuracy of both methods for different bit-widths. We observe that the performance of the post-processing quantization method is only slightly worse than the STE for 8 bits to 5 bits. However, the performance drops rapidly when using fewer than 4 bits. Quantization using the STE on the other hand preserves the performance even for binary weight (1-bit) networks. We observe only a slight drop in performance of around $0.9\%$ when going from 8 bit to 1 bit weights using the STE.

As we have seen, weight quantization using the STE can be employed successfully without any substantial loss in accuracy. Also, note that the training overhead introduced by the STE is only marginal. Training with the STE does not take much longer than conventional training. Post-processing quantization performed worse by comparison but has some advantages. It is easy to implement and it can be performed on arbitrary models without changing the model. 

\subsection{Varying Number of MFCC Features}
\label{sec:mfcc_features}
For the previous experiments we kept the number of MFCC features at 10 MFCCs. We have seen that using 10 MFCC features is sufficient for achieving a reasonable performance. Here, we aim to explore if using more than 10 MFCC features affects the performance, number of operations and memory requirements of the networks.

To compare performances we run 4 NAS experiments using 10, 20, 30 and 40 MFCC respectively. As before we select $\beta \in \{0, 1, 2, 4, 8, 16\}$ and search for optimal architectures followed by retraining of the weights. The channel multiplier was kept at $\omega = 1$. In order to compare to the baseline, weight quantization to 8 bits using the STE is performed during training.

\begin{figure}[t!]
	\centerline{\includegraphics[width=0.85\linewidth]{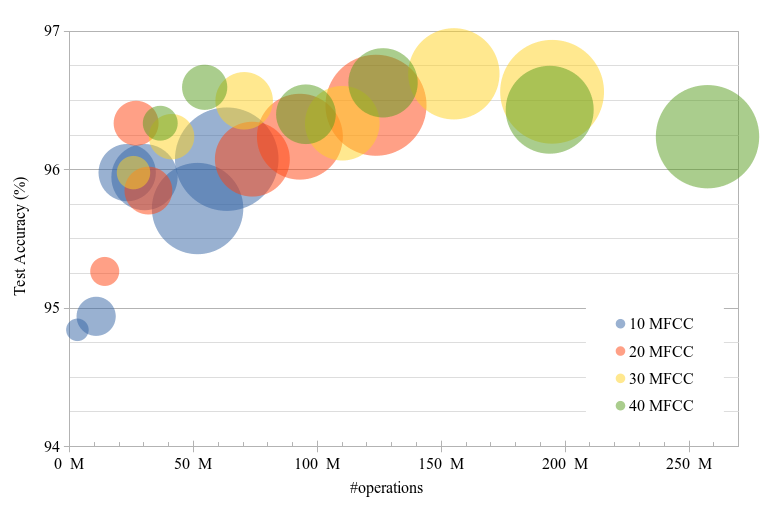}}
	\caption{Test accuracy versus number of operations using Multi-objective NAS with varying trade-offs $\beta$ for 10 (blue), 20 (red), 30 (yellow) and 40 (green) MFCCs. The model size corresponds to the circle size.}
	\label{fig:mfcc_test}
\end{figure}

Figure~\ref{fig:mfcc_test} shows the results for using 10, 20, 30 and 40 MFCCs for several trade-offs $\beta$. We also provide a comparison of models found by NAS with $\beta=8$ using 10, 20, 30 and 40 MFCCs in Table~\ref{tab:mfcc_result_comparison}. 

\begin{table}[t!]
	\caption{Test accuracy, number of operations and memory requirements (for storing weights) of models obtained by NAS with $\beta=8$ using a different number of MFCCs as features.}
	\begin{center}
		\begin{tabular}{|c|c|c|c|c|}
			\hline
			\# MFCCs & Test Accuracy (\%) & Operations & Memory \\\hline\hline
			10 & 94.9 & 10.9 M & 258.8 kB \\\hline
			20 & 96.3 & 27.1 M & 340.1 kB \\\hline
			30 & 96.2 & 41.4 M & 348.2 kB \\\hline
			40 & 96.6 & 54.7 M & 344.7 kB \\\hline
		\end{tabular}
	\end{center}
	\label{tab:mfcc_result_comparison}
\end{table}

We notice a definite improvement of using 20 MFCCs over 10 MFCCs. In particular, one model found by NAS uses 340.1 kB of memory, 27.1 million operations and achieves 96.3 \% accuracy which outperforms \cite{Zhang2017} by a large margin. We include the model structure of this particular network in Figure~\ref{fig:models}d. Using 30 MFCCs, we obtained the best performing network with a test accuracy of 96.7\% using 1403.3 kB of memory and 155.3 million operations.

\section{Conclusion}
\label{sec:conclusion}
Resource efficient DNNs are the key components in modern keyword spotting (KWS) systems. We show that neural architecture search (NAS) can be used to obtain efficient convolutional neural networks (CNNs) without compromising classification accuracy. The CNN models obtained by NAS achieve state-of-the-art performance while being small enough to fit inside the memory of devices with limited resources such as a microcontroller. To make our results comparable, we performed NAS on the Google speech commands dataset. By using an accuracy/operations trade-off objective for NAS we obtained different models and compared them in terms of accuracy, memory and number of operations. We furthermore showed that weight quantization is a viable option to reduce the memory footprint for storing the CNN weights even further. Even 1 bit weights are sufficient for CNNs without severely degrading the accuracy. We also showed that changing the number of MFCC features can have a substantial impact on the performance of the models.

\section*{Acknowledgement}
This work was supported by the Austrian Science Fund (FWF) under the project number I2706-N31.

\bibliographystyle{IEEEtran}
\bibliography{IEEEabrv,mybibfile}

\end{document}